\documentstyle[aps,multicol,prb,epsfig,floats]{revtex}

\begin{document}

\twocolumn[\hsize\textwidth\columnwidth\hsize\csname@twocolumnfalse%
\endcsname

\draft

\title{Aging and memory effects in $\beta$-hydrochinone-clathrate}

\author{A.V.~Kityk$^{1,2}$, M.C.~Rheinst\"adter$^1$, and K.~Knorr$^1$}
\address{$^1$ Technische Physik, Universit\"at des Saarlandes, 
         66041 Saarbr\"ucken, Germany\\
         $^2$ Institute for Computer Science, Faculty of Electrical Eng.,
              Czestochowa Technical University, 42-200 Czestochowa, Poland}

\author{H.~Rieger}
\address{Theoretische Physik, Universit\"at des Saarlandes, 
         66041 Saarbr\"ucken, Germany}

\maketitle

\date{\today}

\begin{abstract}
  The out-of-equilibrium low-frequency complex susceptibility of the
  orientational glass methanol(73\%)-$\beta$-hydrochinone-clathrate
  is studied using temperature-stop protocols in aging experiments .
  Although the material does not have a sharp glass transition aging
  effects including rejuvenation and memory are found at low
  temperatures. However, they turn out to be much weaker, however,
  than in conventional magnetic spin glasses.
\end{abstract}



]

\section{Introduction}

Disordered and frustrated systems have attracted considerable
attention during the past decade. Much effort in the recent years has
been devoted to investigations of slow dynamics in these systems.
Structural glasses, magnetic spin glasses and their electrical
pendant, orientational glasses are the typical examples of
out-of-equilibrium systems. The most remarkable and common peculiarity
of these materials is physical aging as has been extensively studied
in spin glasses \cite{sg-review} and also observed in orientational
glasses (for instance K$_{0.989}$Li$_{0.011}$TaO$_3$ \cite{{ref2,ref3}})
and disordered ferroelectrics (for instance
KTa$_{0.973}$Nb$_{0.027}$O$_3$ \cite{doussineau}). 

Recent experiments on the insulating (Heisenberg-like) spin glass
CdCr$_{1.7}$In$_{0.3}$S$_4$, showed that spin glasses can even
memorize some of the features of the way in which the system has been
prepared \cite{sg-mem}: Prolonged aging at a constant temperature
$T_1$ during slow cooling, a so-called temperature-stop, causes a dip
around this temperature in the temperature dependence of the
imaginary part of the magnetic susceptibility upon reheating. Even
multiple aging temperatures $T_i$ can be memorized in a corresponding
multi-temperature-stop experiment \cite{multi-stop}.  Other materials,
like an Ising spin glass Fe$_{0.5}$Mn$_{0.5}$TiO$_3$ \cite{dupuis}, an
interacting Fe-C nanoparticle system \cite{jonsson} and the disordered
ferroelectrics Pb(Mg$_{1/3}$Nb$_{2/3}$)O$_3$ \cite{colla} and
KTa$_{0.973}$Nb$_{0.027}$O$_3$ \cite{doussineau} displayed similar but
less pronounced memory features in these temperature stop experiments.

Whereas many aging features of disordered materials in general and of
spin glasses in particular are theoretically quite well understood on
a phenomenological and on the mean-field level \cite{theory}, the
theoretical status for these memory experiments is less clear.  Two
ingredients appear to be important: 1) A pronounced sensibility of the
spin glass state with respect to temperature changes to explain
so-called rejunvenation effects when the temperature is changed during
aging (sometimes also called {\it chaos} in spin glasses
\cite{chaos}), and 2) strongly coupled regions in equilibrated domains
whose once developped correlations are hard to destroy when the
temperature is changed and which can serve as a nucleation centers
when the temperature is raised again to explain the memory effect.
Obviously such features are absent in for instance pure ferromagnets,
which neither show rejuvenation effects when the temperature is
changed during domain growth nor memory effects. However, strongly
disordered ferromagnets display a very weak form of memory which is
commonly attributed to the reconformation of domain walls upon
temperature changes \cite{ferro-mem}. Loosely speaking one might
summarize the current status of the theoretical picture as: The walls
are the glassy objects in random ferromagnets, whereas spin glass
domains behave glassy as a whole, in particular because they are most
probably fractal objects \cite{fractal} and surface (wall) as well as
bulk contributions to aging effects are comparable. Thus is varying
strength in which the materials mentioned above display the chaos and
memory effect in aging experiments can be interpreted as an indication
for the degree in which aging can actually be described by a simple
domain growth like in random ferromagnets or random field systems, see
for instance the discussion in \onlinecite{ref2}.

From this viewpoint we discuss in this paper the electric dipole
``pseudospin'' glass system
methanol$(x=73\%)$-$\beta$-hydrochinone-clathrate. It does not have a
sharp transition temperature, glassy features become simply more and
more pronounced as the temperature is decreased, similar to
two-dimensional magnetic spin glasses \cite{2dsg}, which shows
strong aging effects at low temperature but do not have a 
spin glass transition. We demonstrate that the clathrate we study
even displays at low temperatures a rejuvenation and memory
effect in one-temperature-stop experiments and also a weaker effect in
two-temperature-stop experiment. The paper is organized as follows: In
section II we discuss some of the physical properties of the
methanol-clathrate that we studied, section III deals with a number of
experimental details.  In section IV we present our results on
temperature-stop aging experiments and in section C we discuss our
findings on the background of the current theoretical understanding of
aging in disordered and glassy systems.

\section{The Material}

In this paper we present results on the aging and memory phenomena
observed in the low frequency ac-permittivity of dipolar glass system
methanol$(x=73\%)$-$\beta$-hydrochinone-clathrate. In the
$\beta$-modification the quinol (HO-C$_6$H$_4$-OH) molecules form a
H-bonded rhombohedral R$\overline{3}$ lattice with nearly spherical
cavities, one per unit cell \cite{disorder}. The methanol guest
molecules residing in these cavities are bound to the quinol host
lattice by weak dispersion forces, only. Therefore, they can reorient
relatively freely. At 65 K the methanol$(97 \%)$-clathrate shows an
antiferroelectric transition \cite{woll}. The structure consists
basically of ferroelectric chains running along the hexagonal axis
which are arranged in sheets of alternating sign. Similar phase
transitions have been observed also for other polar guest molecules.
Matsuo and Suga \cite{matsuo} have found the transition temperature
$T_s$ to scale with square of the dipole moment $\mu$ of the free
molecules. It thus suggests that the dominant coupling between the
guest molecules is the electrostatic dipole-dipole interaction. Higher
concentrated samples ($x>x_c \approx 0.76)$ of
methanol$(x)$-$\beta$-hydrochinone-clathrate show conventional
ordering via a first-order phase transition, whereas the others
($x<x_c$) freeze into dipole glasses.

Woll {\it{et al}} \cite{woll} have studied the dielectric relaxation
as a function of temperature, frequency and methanol concentration.
The static dielectric constant along hexagonal $c$-axis
$\varepsilon_{cs}$ shows strong deviation from the Curie-Weiss
behavior already at 250 K. The temperature dependence of
$\varepsilon_{cs}$ at higher temperatures is well described within
mean-field approximation by the one-dimensional (1D) Ising model with
a coupling $J_c$ to nearest neighbors within the chain and an
inter-chain interaction with coupling constant $J_{\perp}$. The
coupling constants are consistent with the electrostatic dipole-dipole
interaction, the elementary dipole moments being practically identical
with that of the free methanol molecule. At lower $T$,
$\varepsilon_{cs}(T)$ deviates from the Ising-behavior and finally
decreases with decreasing $T$, indicating the onset of 3D
antiferroelectric inter-chain correlations.  Such a behavior persists
also for samples with lower concentrations of methanol: the static
dielectric constant possesses a broad maximum in the region of about
55 K. For samples with a higher concentration, 3D-correlations
eventually lead to the antiferroelectric phase transition. For samples
with lower concentration the antiferroelectric fluctuations presumably
prevent total ferroelectric ordering.  Accordingly, in the
low-temperature region one can expect the coexistence of different
types of locally ordered regions, i.e. clusters of polar and antipolar
types. The dielectric relaxation therefore depends strongly on the
dynamics of polarization clusters in an ac-field.

\begin{figure}
\epsfig{file=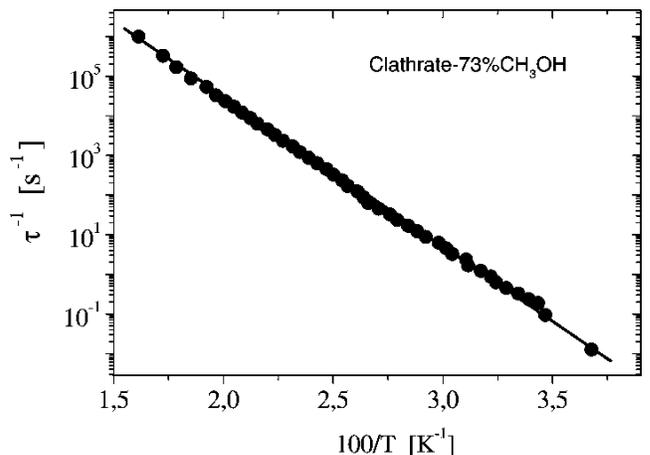, angle=-90, width=0.97\linewidth}
\caption{Inverse average relaxation time
  $\tau^{-1}$ of methanol ($73\%$)-clathrate as a function of $1/T$.}
\label{tau}
\end{figure}

As it was shown before\cite{woll}, the relaxation processes are
characterized by rather broad distribution of relaxation times $\tau$,
which is typical for disordered systems. In Fig.~\ref{tau} the
inverse average relaxation time is plotted against the inverse
temperature in a log-linear plot. The data lay on a straight line
which implies an Arrhenius law
\begin{equation}
\tau^{-1}\sim f_0\exp(-E_A/T)
\label{arr}
\end{equation}
rather than a Vogel-Fulcher behavior $\tau^{-1}\sim
f_0\exp(-E_A/(T-T_{vf}))$ as in structural glasses \cite{donth} or
critical slowing down $\tau^{-1}\sim(T-T_c)^{z\nu}$ in spin glasses
\cite{binder}. In spin glasses the critical temperature $T_c$ at which
the relaxation time diverges is identical to the spin glass transition
temperature at which the non-linear susceptibility and hence the spin
glass correlation length diverges. In structural glasses the
Vogel-Fulcher temperature $T_{vf}$ can serve as a lower bound for the
strongly cooling rate dependent concept of a glass transition
temperature.  Fig.\ \ref{tau} tells us that our methanol
(73\%)-clathrate does {\it not} show a well defined glass transition
in the temperature range 25-100K, the dynamics simply gets slower with
decreasing temperature according to (\ref{arr}). It therefore differs
strongly from other popular examples of dipolar glasses, e.g.
Rb$_{1-x}$(NH$_4$)$_x$H$_2$PO$_4$ \cite{pirc} (member of KDP family)
or cubic perovskites, such as K$_{1-x}$Li$_x$TaO$_3$ \cite{kleemann}.
Nevertheless the freezing of the dipole moments of the 73\% clathrate
is to some part a collective process. Not only the crystal field of
the cavity, but also the interaction between the dipoles contribute to
the Arrhenius barrier $E_A$ in (\ref{arr}). As we will demonstrate in
this paper, aging, rejuvenation and memory phenomena are observable in
the aformentioned temperature range.

\section{Experimental details}

The single crystal of methanol(73\%)-clathrate has been grown from a
saturated solution of quinol, methanol and n-propanol at 313 K. The
sample was prepared in the form of a thin parallel plate ($d \approx
0.5$ mm). The faces were oriented perpendicular to the hexagonal
$c$-axis. Gold films deposited on these faces serve as electrodes. 
The real and imaginary part of the dielectric constant
$\varepsilon^*(\omega)=\varepsilon'(\omega)+i \varepsilon''(\omega)$
are directly related to the capacitance and loss. The geometrical
capacitance as calculated from the thickness and electrode area 
was 0.3pF

The complex dielectric constant was measured with a frequency
response analyzer "Solatron" SI-1255 covering a frequency band from
0.1 mHz to 10 MHz. The amplitude of ac-field was about 50 V/cm.
Measurements of the electric polarization $P$ as function of the
electric field $E$ along the hexagonal axis show perfect linear
behavior up to the maximum field of about 40 kV/cm. It assures that
the measured susceptibility clearly appears in the frame of the linear
response. The samples were placed into a He-flow cryostat with a stability 
of the temperature control of about 0.002 K.

\begin{figure}
\epsfig{file=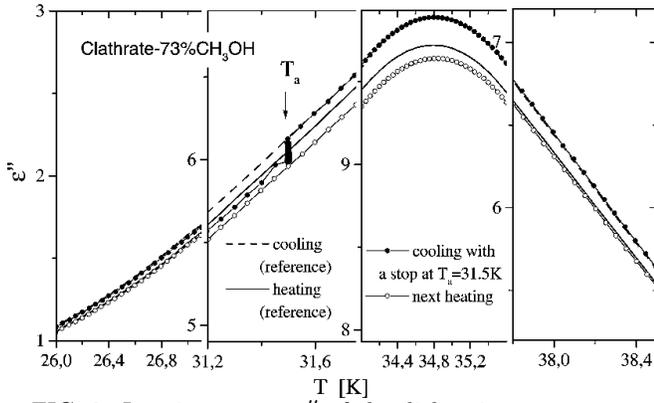, angle=-90, width=1.0\linewidth}
\caption{Imaginary part $\varepsilon''$ of the
  dielectric constant measured at 2 Hz in methanol(73\%)-clathrate.
  Shown are the reference curves $\varepsilon''_{ref}(T)$ for
  continuous cooling and heating with $|dT/dt|$ of about $0.1$ K/min
  and the data for the single temperature-stop (at $T_a=31.5K$ for
  about 14h) aging experiment.} 
\label{eps-im}
\end{figure}

\section{Temperature step experiments}

Motivated by recent aging and memory experiments on spin glasses
\cite{sg-mem} we performed aging and memory experiments on
methanol(73\%)-clathrate using the following temperature protocol.
First reference curves are recorded as a function of temperature at
rather slow cooling and heating ($|dT/dt|$=0.1-0.3 K/min). In the
aging experiment the sample is then continuously cooled with the same
speed, but is additionally kept at a certain constant temperature
$T_a$ for a waiting time of several hours. The ac-susceptibility
$\chi$ relaxes downwards (aging effect) but upon a subsequent cooling
it approaches its reference value at somewhat lower temperature
$T=T_a- \Delta T$ (rejuvenation).  During the next re-heating the
temperature dependence $\chi(T)$ shows a broad dip in the vicinity of
$T_a$, which, of course, does not exist in the reference curve.
Accordingly, the system "remembers" its aging temperature $T_a$
(memory effect). This anomaly completely disappear after overheating
for a few degrees above $T_a$, so subsequent temperature cycling
around $T_a$ shows again the reference behavior.

Fig.~\ref{eps-im} shows such a sequence of several cycling
experiments. The results are presented only for a few temperature
intervals, which have different vertical scales. This unusual
presentation is necessary in order to make the smlaa differences at
the various traces visible. The reference behavior of $\varepsilon''(T)$
are first recorded as a function of temperature at cooling/heating
rate $|dT/dt|$ of about $0.1$ K/min. One can see that the reference curve
$\varepsilon''_{ref}(T)$ at heating is always somewhat
lower than at cooling thus clearly indicating on an
out-of-equilibrium behavior. However, if cooling is interrupted for a
time $t_a$ (in our experiment $t_a \approx$ 50 ks) at constant
temperature (e.g. $T_a=31.5$ K) the imaginary part $\varepsilon''$
decreases by the effect of aging. Upon a subsequent cooling
$\varepsilon''(T)$ approaches its reference value
$\varepsilon''_{ref}$ at somewhat lower temperature of about $T
\approx 25 K$. During the next re-heating the temperature dependence
$\varepsilon''(T)$ shows a little bit lower value than reference. This
difference again disappears at subsequent cooling-heating cycles.

\begin{figure}
\epsfig{file=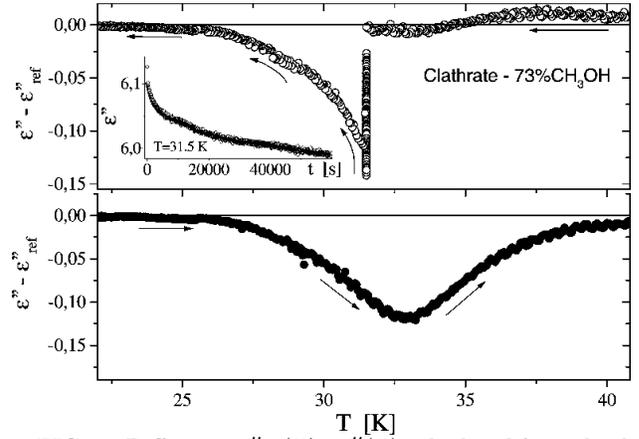, angle=-90, width=0.95\linewidth}
\caption{Difference
  $\varepsilon''_{ref}(T)-\varepsilon''(T)$ calculated from the data
  presented in Fig.\ref{eps-im}. Insert shows the long-time relaxation
  of $\varepsilon''$ recorded at $T=T_a=31.5$K }
\label{eps-diff}
\end{figure}

\begin{figure}
\epsfig{file=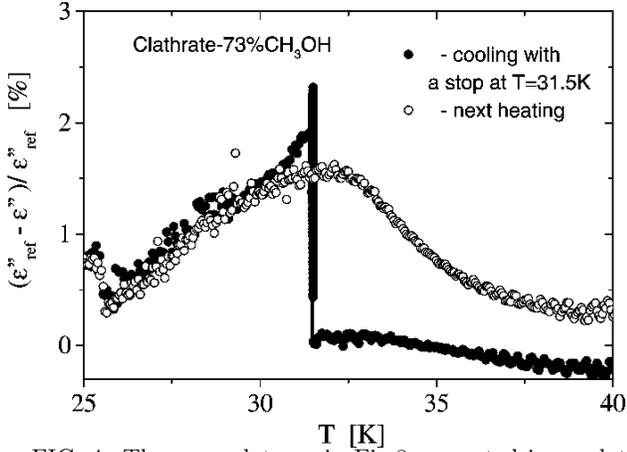, angle=-90, width=0.95\linewidth}
\caption{The same data as in Fig.8 presented in a
  relative form (i.e.
  $e=(\varepsilon''_{ref}(T)-\varepsilon''(T))/\varepsilon''_{ref}(T)$
  vs the temperature).} \label{eps-rel}
\end{figure}

In order to highlight the consequences of isothermal aging
(Fig.~\ref{eps-im}) it is quite convenient to present the data as
difference between the curves recorded during the second and the first
(reference) cooling-heating circles (see Fig.~\ref{eps-diff}). One can
clearly see the sequence of the several effects: aging at $T=T_a$ with
long-time evolution (see insert in Fig.~\ref{eps-diff}), the
rejuvenation upon the subsequent cooling and the memory effect at
further re-heating leading to the broad dip centered close to the
aging temperature $T_a$. Quite similar behavior has been recently
reported in disordered ferroelectric KTa$_{0.973}$Nb$_{0.027}$O$_3$
\cite{doussineau}; the authors have presented their results in the
same manner as in the Fig.~\ref{eps-diff}. In fact, the data presented
in differential form are not able to give true impression about the
aging behavior. For example, the rejuvenation observed at the low
temperatures in methanol-clathrate could be simply due to the almost
complete dynamical freezing. The reference curve
$\varepsilon''_{ref}(T)$ approaches in this temperature range its zero
level and since $\varepsilon''_{ref}(T)>\varepsilon''(T)$ the
difference $\varepsilon''_{ref}(T)-\varepsilon''(T)$ will always have
the same tendency. In order to avoid this problem one must present the
data in a relative form, i.e. as
$e(T)=(\varepsilon''_{ref}(T)-\varepsilon''(T))/\varepsilon''_{ref}(T)$
vs. the temperature.  The corresponding dependence is shown in
Fig.~\ref{eps-rel}.

\begin{figure}
\epsfig{file=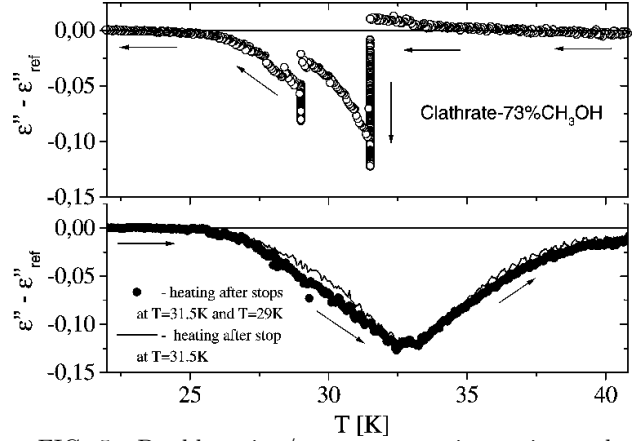, angle=-90, width=0.95\linewidth}
\caption{Double aging/memory experiment in
  methanol (73\%)-clathrate. The dependence
  $\varepsilon''_{ref}(T)-\varepsilon''(T)$ vs. the temperature is
  shown for cooling and heating runs. The experimental procedure is
  similar to this, which is described in Fig.7 with the difference
  only that the second cooling run was twice interrupted at the
  temperatures $T_{a1}=31.5$ K and $T_{a2}=29$ K for the time of about
  50 ks in each case.} \label{double-diff}
\end{figure}

\begin{figure}
\epsfig{file=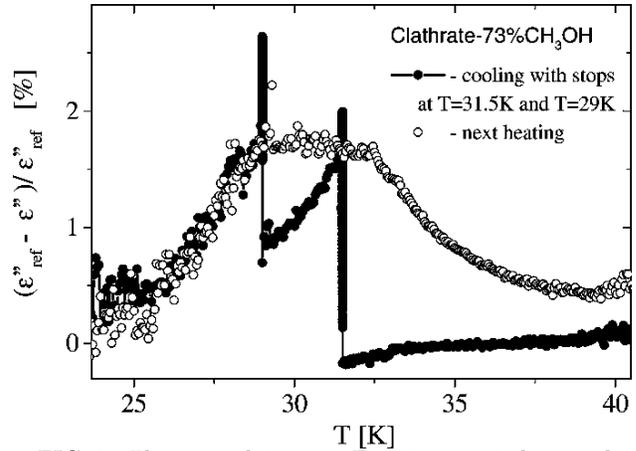, angle=-90, width=0.95\linewidth}
\caption{The same data as in Fig. \ref{double-diff} presented in a
  relative form (i.e.  $e=(\varepsilon''_{ref}(T)-\varepsilon''(T))
  /\varepsilon''_{ref}(T)$ vs. the temperature).} \label{double-rel}
\end{figure}

Fig.~\ref{double-diff} and Fig.~\ref{double-rel} present so-called double-
aging/memory experiment: the cooling process was interrupted twice at
the temperatures $T_{a1}=31.5$ K and $T_{a2}=29$ K for a time
$t_{a1}=t_{a2}=50$ ks. For both temperatures the imaginary part
$\varepsilon''$ decreases by the effect of aging. Upon a subsequent
cooling one can see rejuvenation, whereas during the re-heating the
temperature dependence $\varepsilon''_{ref}(T)-\varepsilon''(T)$ shows
again a broad maximum. Two anomalies which are expected to occur due
to the memory effect are not well separated in our case. The
temperature interval ($\Delta T=T_{a1}-T_{a2}= 2.5$ K) is too small
for that, and additionally, the aging and memory at the lower
temperature $T_{a2}$ are characterized by much weaker anomalies.
Nevertheless, one can see some difference in the temperature
dependence $\varepsilon''_{ref}(T)-\varepsilon''(T)$ obtained during
the heating in the single-aging(see Fig.~\ref{double-diff}, line) and
double-aging (points) experiments. Both these effects are much more pronounced when presented as relative changes in Fig.~\ref{double-rel}. We
note that similar observation of the double-aging/memory effects was
recently reported in \cite{doussineau} for
KTa$_{0.973}$Nb$_{0.027}$O$_3$ crystals.  By choosing of a larger
temperature interval between the aging temperatures (of about 5 K),
authors obtained here two well separated dips corresponding to the
memory effect.


\section{Discussion}

Since spin glasses, in contrast to structural glasses, have an order
parameter, the EA-order parameter, it appears natural to interpret
aging phenomena at low temperatures in terms of a slowly increasing
domain length --- indications for this have been found numerically in
microscopic SG models \cite{corr-num} as well in recent experiments
\cite{corr-exp}. A simple domain growth, however, cannot explain the
memory effect as evidenced in these multiple-temperature-stop
experiments: in spin glasses domains at different temperatures have to
be uncorrelated to some extent, in contrast to domains in a
ferromagnet for instance. This feature is often called {\it chaos} in
reminiscence of the notion of a finite overlap length of {\it
  equilibrium} states in finite-dimensional spin glasses \cite{chaos}.
In addition to this temperature-chaos in spin glasses to explain the
memory effect the once grown domains should also not be completely
destroyed when the temperature is lowered but should retain one or
several nuclei at least for times comparable to the aging time at
lower temperatures --- such an idea was put forward recently in
\cite{multi-stop}.  

Comparing the outcome of our experiments, in particular the
two-temperature-stop experiments depicted in Fig. \ref{double-diff}
and \ref{double-rel} with the corresponding experiments on the
insulating spin glass CdCr$_{1.7}$In$_{0.3}$S$_4$ \cite{sg-mem}, shows
that the aforementioned spin glass features, including rejuvenation
and memory, are much less pronounced. An obvious difference is of
course that CdCr$_{1.7}$In$_{0.3}$S$_4$ has a clear phase transition
at a temperature $T_g=16.7K$ to a spin glass phase at (and below)
which any relaxation time of the system diverges and below which the
multiple-temperature-stop experiments have been performed. On the
other hand, no indication of a sharp transition can be found in our
clathrate although the relaxation time below 20K already exceeds the
experimental time-scales accessible to us. Quite frequently aging
properties of spin glasses at very low temperatures that do {\it not}
have a phase transition at finite temperatures --- like pseudo-2$d$
materials or thin films \cite{2dsg} --- resemble those of real spin
glasses (i.e.\ those with a clear phase transition) in the frozen
phase. It might well be that this is not so for chaos and memory, for
which one actually needs to have in a spin glass phase.  Another
explanation of the difference could be that our orientation glass, the
methanol clathrate we studied here, is less glassy than expected from
other experiments \cite{woll} and resembles more the disordered
dielectric K$_{1-x}$Li$_x$TaO$_3$: Here aging effects can be
interpreted as a simple domain growth of varying speed and the system
appears to be more closely related to a random field system rather
than a spin glass. Certainly it would be worthwhile to scrutinize the
details of the microscopic mechanism underlying the aging effects we
presented in a more detailed form.

{\bf Acknowledgements:} This work has been partly supported by the
Deutsche Forschungsgemeinschaft (SFB 277).

\end{document}